\documentclass[a4paper,11pt]{article}
\pdfoutput=1 

\usepackage{jheppub} 

\usepackage[T1]{fontenc} 
\usepackage{amssymb,amsmath,bm,natbib}
\usepackage{color}
\usepackage{slashed}
\usepackage{graphics}
\usepackage{graphicx}
\usepackage[utf8]{inputenc}
\usepackage[caption=false]{subfig}
\usepackage{hyperref}
\usepackage{url}
\usepackage{dsfont}
\usepackage{float}
\usepackage{cancel}
\usepackage{units}
\usepackage{blindtext}
\usepackage[utf8]{inputenc}
\usepackage{upgreek}
\usepackage{booktabs}

\preprint{CERN-TH-2020-142, PSI-PR-20-14, ZU-TH 29/20}

\title{Implications of $\textit{SU}(2)_L$ gauge invariance for constraints on Lorentz violation}

\author[a,b,c]{Andreas Crivellin,}
\author[b]{Fiona Kirk,}
\author[d]{and Marco Schreck}

\affiliation[a]{CERN Theory Division, CH--1211 Geneva 23, Switzerland}
\affiliation[b]{Paul Scherrer Institut, CH--5232 Villigen PSI, Switzerland}
\affiliation[c]{Physik-Institut, Universit\"at Z\"urich, Winterthurerstra\ss{}e 190, CH--8057 Z\"urich, Switzerland}
\affiliation[d]{Departamento de F\'{i}sica, Universidade Federal do Maranh\~{a}o, Campus Universit\'{a}rio do Bacanga, S\~{a}o Lu\'{i}s -- MA, 65085-580, Brazil}

\emailAdd{andreas.crivellin@cern.ch}
\emailAdd{fiona.kirk@psi.ch}
\emailAdd{marco.schreck@ufma.br}

\abstract{Lorentz invariance may only be broken far above the electroweak scale, since violations are experimentally stringently constrained. Therefore, the Standard-Model Extension parameterizing Lorentz violation (LV) via (higher-dimensional) field theory operators is manifestly $\mathit{SU}(2)_L$ gauge-invariant. As a consequence, LV in neutrinos implies LV in charged leptons and vice versa. This allows us to obtain estimated sensitivities for flavour-changing operators in the charged-lepton sector from neutrino oscillations as well as sensitivities for flavour-diagonal neutrino effects from high-precision electron experiments. We also apply this method to an analysis of time-of-flight data for neutrinos (detected by IceCube) and photons from gamma ray bursts where discrepancies have been observed. Our conclusion is that an explanation of the arrival time difference between neutrino and photon events by dim-5 operators in the neutrino sector would lead to unacceptably large LV effects in the charged-lepton sector.}


\def\ring#1{{\mathaccent'27 #1}}

\begin{document}
\maketitle
\flushbottom

\section{Introduction}

Lorentz invariance is the cornerstone that both the Standard Model (SM) of elementary particle physics and General Relativity rest on. However, underlying theories such as strings~\cite{Kostelecky:1988zi,Kostelecky:1991ak,Kostelecky:1994rn} or loop quantum gravity~\cite{Gambini:1998it,Bojowald:2004bb} as well as models that exhibit small-scale spacetime structures~\cite{AmelinoCamelia:1999pm,Carroll:2001ws,Klinkhamer:2003ec,Bernadotte:2006ya,Hossenfelder:2014hha} could result in violations of this symmetry at very high energies (e.g., the Planck scale). Since, in general, a violation of Lorentz symmetry leads to modified particle properties such as energy-dependent and/or direction-dependent dispersion relations and field equations (see, e.g., refs.~\cite{AmelinoCamelia:1997gz,Coleman:1997xq,Coleman:1998ti,Aloisio:2000cm}) its effects can be observable at energies far below the Planck energy where Earth-based experiments or astrophysical observations are performed. Clearly, a detection of LV would arguably be the most astounding discovery in fundamental physics since the establishment of quantum mechanics and relativity around one century ago.

Deviations from Lorentz invariance are commonly quantified in the effective field theory framework called the Standard-Model Extension (SME)~\cite{Colladay:1996iz,Colladay:1998fq,Kostelecky:2003fs}. Within the SME (in the absence of gravity), LV is described by background fields in spacetime that arise as vacuum expectation values of tensor-valued fields in a fundamental theory. The latter are nondynamical and are contracted with field operators in such a way that coordinate invariance is maintained. The background fields give rise to preferred spacetime directions and the strength of LV is parameterized by controlling coefficients. Importantly, since LV is assumed to originate from very high energies, the operators of the SME are manifestly $\mathit{SU}(2)_L$ gauge-invariant. Since a violation of {\em CPT} invariance implies LV within effective field theory~\cite{Greenberg:2002uu}, both types of violations are connected to each other and all {\em CPT}-violating operators, which are coordinate-invariant, are automatically contained in the SME.

A multitude of tests of Lorentz invariance have been carried out over the past decades. These range from table-top precision experiments with atomic clocks to astrophysical observations of ultra-high-energy cosmic rays, photons, and neutrinos (see, e.g., refs.~\cite{Mattingly:2005re,Liberati:2013xla,Tasson:2014dfa} for an overview). As a conclusive signal of LV has not been found so far, the experiments have led to constraints on controlling coefficients of the effective operators of the SME, which are compiled in (yearly updated) data tables~\cite{Kostelecky:2008ts}.

Even though the SME is $\mathit{SU}(2)_L$-invariant (like the Lorentz-invariant SM Effective Theory~\cite{Buchmuller:1985jz}), the bounds of ref.~\cite{Kostelecky:2008ts} are given in the broken theory in which, e.g., left-handed charged leptons and neutrinos are independent fields. In this article, we study how different experimental limits on LV are related via $\mathit{SU}(2)_L$ invariance. We correlate the charged-lepton sector to the neutrino sector and vice versa and show, in particular, that one can infer constraints on LV for neutrinos from associated constraints in the charged-lepton sector. This approach leads to potential sensitivities based on existing experimental limits. For that purpose, we will start with the minimal SME that involves operators of mass dim-3 and 4 only and, subsequently, include dim-5 operators. This consideration will demonstrate that LV in the neutrino sector, as studied in the context of the former OPERA excess~\cite{Adam:2011faa} and recently deduced from data on IceCube neutrino events in refs.~\cite{Huang:2018ham,Huang:2019etr}, would imply LV for charged leptons that clashes with existing experimental constraints.

The paper is organized as follows. Section~\ref{sec:sme-lepton-sector} gives a brief introduction to the SME fermion sector and states the properties that are essential for our analysis. In section~\ref{sec:connection-coefficients} we apply the argument based on $\mathit{SU}(2)_L$ invariance to coefficients of the minimal neutrino and charged-lepton sector. Section~\ref{sec:consequences-neutrino-analyses} presents the implications of this procedure for the aforementioned analysis of time-of-flight data of ultra-high-energy IceCube neutrinos. Finally, we conclude in section~\ref{sec:conclusions}. Details that are worthwhile mentioning, but not necessary to understand the contents of the main body of the paper, will be relegated to appendices~\ref{sec:derivation-constraints-details}, \ref{sec:fit-neutrino-time-of-flight}. Natural units with $\hbar=c=1$ will be used unless otherwise stated.

\section{Lepton sector of the SME}
\label{sec:sme-lepton-sector}

As motivated in the introduction, LV is usually assumed to occur at very high energies. The effective framework valid at low energies, the SME, includes LV operators of mass dim-3 and 4, classified in ref.~\cite{Colladay:1998fq} for all particle sectors but gravity, which is considered in ref.~\cite{Kostelecky:2003fs}. Operators of mass dimensions larger than 4 can be found in ref.~\cite{Kostelecky:2009zp} for photons, in ref.~\cite{Kostelecky:2011gq} for neutrinos, and in ref.~\cite{Kostelecky:2013rta} for Dirac fermions. Respecting $\mathit{SU}(3)_c\times \mathit{SU}(2)_L\times\mathit{U}(1)_Y$, we have the following Lorentz-violating modification of the lepton sector:\footnote{In order to have local invariance under $\mathit{SU}(3)_c\times \mathit{SU}(2)_L\times\mathit{U}(1)_Y$, the partial derivatives in the Lagrange density need to be suitably promoted to covariant derivatives~\cite{Ding:2016lwt,Kostelecky:2020hbb}. Our argument relies on global gauge invariance in the lepton sector, i.e., we consider the free Lagrangian and neglect all Lorentz-violating terms in interactions and the pure gauge sector.}\footnote{Note the connection to $\mathcal{L}^{(d)}_{\mathrm{lepton},D}$ of table~XVII in the recent paper of ref.~\cite{Kostelecky:2020hbb} that complements the minimal (gravitational) SME by including higher-derivative contributions.}
\begin{align}
\mathcal{L}&=\frac{1}{2}\sum_{\Psi} \overline{\Psi}_A[(\hat{c}_\Psi)^{\mu\nu}_{AB}\mathrm{i}\partial_{\nu}-(\hat{a}_\Psi)^{\mu}_{AB}]\gamma_{\mu}\Psi_B+\text{H.c.}
\label{eq:lagrangian-sme-lepton-sector}
\end{align}
with $\Psi\in \{L,R\}$ where $L$ $(R)$ is the left-handed (right-handed) lepton $\mathit{SU}(2)_L$ doublet (singlet)
\begin{align}
	L_A&=\begin{pmatrix}
	\nu_A \\
	\ell_A \\
	\end{pmatrix}_L\,,\quad R_A=(\ell_A)_R\,.
\end{align}
The subscripts $L$, $R$ in the latter label chirality. Flavour indices are denoted by capital Latin letters (e.g., $A\in \{\mathrm{e},\upmu,\uptau\}$). Furthermore, $(\hat{a}_{\Psi})^{\mu}_{AB}$ and $(\hat{c}_{\Psi})^{\mu\nu}_{AB}$ are understood as generalizations of the LV coefficients within the minimal SME~ \cite{Colladay:1996iz,Colladay:1998fq}. They can be written as infinite series involving four-derivatives:
\begin{subequations}
\begin{align}
\label{eq:lv-operator-a}
(\hat{a}_{\Psi})^{\mu}_{AB}&=\sum_{\substack{d\geq 3 \\ d \text{ odd}}} a_{\Psi,AB}^{(d)\mu\alpha_1\dots\alpha_{d-3}}(\mathrm{i}\partial_{\alpha_1})\dots (\mathrm{i}\partial_{\alpha_{d-3}})\,,\\
\label{eq:lv-operator-c}
(\hat{c}_{\Psi})^{\mu\nu}_{AB}&=\sum_{\substack{d\geq 4 \\ d \text{ even}}} c_{\Psi,AB}^{(d)\mu\nu\alpha_1\dots\alpha_{d-4}}(\mathrm{i}\partial_{\alpha_1})\dots (\mathrm{i}\partial_{\alpha_{d-4}})\,.
\end{align}
\end{subequations}
Here, $a_{\Psi,AB}^{(d)\mu\alpha_1\dots\alpha_{d-3}}$ and $c_{\Psi,AB}^{(d)\mu\nu\alpha_1\dots\alpha_{d-4}}$ are controlling coefficients (equivalent to Wilson coefficients) that are associated with field operators of mass dim-$d$. The operator in eq.~(\ref{eq:lv-operator-a}) (eq.~(\ref{eq:lv-operator-c})) is {\em C}-odd ({\em C}-even)~\cite{Kostelecky:2008ts} which implies that the coefficients enter with opposite (same) signs in the dispersion relations of fermions and antifermions (cf.~refs.~\cite{Kostelecky:2000mm,Kostelecky:2013rta}). As the operator has an odd (even) number of Lorentz indices, it is also {\em CPT}-odd ({\em CPT}-even), i.e., it generates (no) {\em CPT} violation. Note that since the neutrino is contained within the lepton doublet, any modification of neutrino properties also affects charged leptons.\footnote{Similar arguments relying on $\mathit{SU}(2)_L$ gauge invariance were also employed in ref.~\cite{Jentschura:2019wsr,Jentschura:2020nfe}, although in the context of LV beyond the SME.}

\section{Connection between neutrino and charged-lepton coefficients}
\label{sec:connection-coefficients}

The Lagrangian of eq.~(\ref{eq:lagrangian-sme-lepton-sector}) gives rise to the following modified field equations for left-handed neutrinos and charged leptons:
\begin{subequations}
\begin{align}
\label{eq:modified-dirac-equation-neutrinos}
0&=\left\{\mathrm{i}\cancel{\partial}\delta_{AB}+\left[(\hat{c}_L)^{\mu\nu}_{AB}\mathrm{i}\partial_{\nu}-(\hat{a}_L)^{\mu}_{AB}\right]\gamma_{\mu}\right\}(\nu_B)_L\,,\\[1ex]
0&=\left\{\mathrm{i}\cancel{\partial}\delta_{AB}+\left[(\hat{c}^\ell)^{\mu\nu}_{AB}\mathrm{i}\partial_{\nu}-(\hat{a}^\ell)^{\mu}_{AB}\right]\gamma_{\mu}\right. \notag \\
&\phantom{{}={}}\left.+\left[(\hat{d}^\ell)^{\mu\nu}_{AB}\mathrm{i}\partial_{\nu}-(\hat{b}^\ell)^{\mu}_{AB}\right]\gamma_5\gamma_{\mu}\right\}\ell_B\,,
\end{align}
\end{subequations}
with
\begin{subequations}
\label{eq:lepton-sector-operators}
\begin{align}
(\hat{a}^\ell)^{\mu}_{AB}&=\frac{1}{2}\left[(\hat{a}_L)^{\mu}_{AB}+(\hat{a}_R)^{\mu}_{AB}\right]\,,\\[1ex]
(\hat{b}^\ell)^{\mu}_{AB}&=\frac{1}{2}\left[(\hat{a}_L)^{\mu}_{AB}-(\hat{a}_R)^{\mu}_{AB}\right]\,,\\[1ex]
(\hat{c}^\ell)^{\mu\nu}_{AB}&=\frac{1}{2}\left[(\hat{c}_L)^{\mu\nu}_{AB}+(\hat{c}_R)^{\mu\nu}_{AB}\right]\,, \\[1ex]
(\hat{d}^\ell)^{\mu\nu}_{AB}&=\frac{1}{2}\left[(\hat{c}_L)^{\mu\nu}_{AB}-(\hat{c}_R)^{\mu\nu}_{AB}\right]\,,
\end{align}
\end{subequations}
where the superscript $\ell$ stands for charged lepton. In the following, we will assume that the mass and the interaction eigenbasis for both charged leptons and neutrinos are identical, which can always be achieved if the latter are massless. Furthermore, we will neglect flavour-violating effects in the charged-lepton sector ($A\neq B$) where the related constraints would be weak due to missing interference with the SM contributions.

According to eq.~(\ref{eq:lepton-sector-operators}), any modification in the neutrino sector implies LV in the charged-lepton sector. The converse only holds when left-handed charged leptons are modified, i.e., in case of nonvanishing operators $(\hat{a}_L)_{AB}^{\mu}$, $(\hat{c}_L)_{AB}^{\mu\nu}$. We will mainly deal with the setting $(\hat{a}_R)_{AB}^{\mu}=(\hat{c}_R)_{AB}^{\mu\nu}=0$. In addition, it is interesting to consider the scenario
$(\hat{a}_L)_{AB}^{\mu}=(\hat{a}_R)_{AB}^{\mu}$, $(\hat{c}_L)_{AB}^{\mu\nu}=(\hat{c}_R)_{AB}^{\mu\nu}$, which can be realized within left-right (LR) symmetric models~\cite{Mohapatra:1974hk} if the LR breaking scale is below the LV scale. In what follows, we will infer new potential sensitivities from already existing ones at the cost of taking a series of assumptions such as the absence of fine-tuned cancelations between different LV coefficients. These new sensitivities show the extend to which current experiments are able to probe physics at the Planck scale possibly affecting the flavour-offdiagonal charged-lepton sector and the flavour-diagonal neutrino sector.
\begin{table}
\centering
\begin{tabular}{|cccc|}
\toprule
$\mu$         & $\mathrm{Re}(a^{\ell})_{\mathrm{e}\upmu}^{\mu}$ & $\mathrm{Re}(a^{\ell})_{\mathrm{e}\uptau}^{\mu}$ & $\mathrm{Re}(a^{\ell})_{\upmu\uptau}^{\mu}$ \\
\midrule
$T$           & $\unit[5\times 10^{-21}]{GeV}$                  & $\unit[5\times 10^{-20}]{GeV}$                   & $\unit[5\times 10^{-25}]{GeV}$ \\
$X$           & $\unit[5\times 10^{-21}]{GeV}$                  & $\unit[5\times 10^{-20}]{GeV}$                   & $\unit[5\times 10^{-24}]{GeV}$ \\
$Y$           & $\unit[5\times 10^{-22}]{GeV}$                  & $\unit[5\times 10^{-20}]{GeV}$                   & $\unit[5\times 10^{-24}]{GeV}$ \\
$Z$           & $\unit[5\times 10^{-20}]{GeV}$                  & $\unit[5\times 10^{-20}]{GeV}$                   & --- \\
\toprule
              & $\mathrm{Im}(a^{\ell})_{\mathrm{e}\upmu}^{\mu}$ & $\mathrm{Im}(a^{\ell})_{\mathrm{e}\uptau}^{\mu}$ & $\mathrm{Im}(a^{\ell})_{\upmu\uptau}^{\mu}$ \\
\midrule
$T$           & $\unit[2\times 10^{-20}]{GeV}^{\dagger}$        & $\unit[5\times 10^{-20}]{GeV}$                   & $\unit[3\times 10^{-24}]{GeV}^{\diamond\dagger}$ \\
$X$           & $\unit[5\times 10^{-21}]{GeV}$                  & $\unit[5\times 10^{-20}]{GeV}$                   & $\unit[5\times 10^{-21}]{GeV}$ \\
$Y$           & $\unit[5\times 10^{-22}]{GeV}$                  & $\unit[5\times 10^{-20}]{GeV}$                   & $\unit[5\times 10^{-21}]{GeV}$ \\
$Z$           & $\unit[5\times 10^{-20}]{GeV}$                  & $\unit[5\times 10^{-20}]{GeV}$                   & --- \\
\toprule
              & $\ring{a}^{\ell}_{\mathrm{e}\upmu}$             & $\ring{a}^{\ell}_{\mathrm{e}\uptau}$             & $\ring{a}^{\ell}_{\upmu\uptau}$ \\
\midrule
              & $\unit[5 \times 10^{-21}]{GeV}$                 & ---                                              & $\unit[5\times 10^{-25}]{GeV}$ \\
\bottomrule
\end{tabular}
\caption{Potential sensitivities of flavour-off-diagonal dim-3 vector-valued coefficients $(a^{\ell})^{\mu}_{AB}$ in the charged-lepton sector obtained from limits in the neutrino sector via $\mathit{SU}(2)_L$ invariance (see appendix~\ref{sec:derivation-constraints-details} for calculational details). All sensitivities are two-sided unless those with the symbol~$\diamond$ that are upper ones. Here we assumed that only left-handed charged-lepton fields are modified, i.e., $(a_R)^{\mu}_{AB}=0$. Then the (unstated) sensitivities on $(b^{\ell})_{AB}^{\mu}$ equal those on $(a^{\ell})_{AB}^{\mu}$. In the LR-symmetric case, the values for $(a^{\ell})_{AB}^{\mu}$ must be multiplied by a factor of~2 while no bounds on $(b^{\ell})_{AB}^{\mu}$ can be obtained. Whenever possible, the sensitivities were inferred from values listed in table~S4 of ref.~\cite{Kostelecky:2008ts} and the symbol~$\dagger$ indicates data used from table~D29 of the same reference. All values are given in the standard Sun-centered inertial reference frame~\cite{Kostelecky:2008ts}. The index $T$ stands for the time component and $\{X,Y,Z\}$ for the spatial components. The notation $\ring{a}^{\ell}_{AB}$, etc. denotes isotropic parts of the coefficients (see section~IV.B in ref.~\cite{Kostelecky:2013rta} for their definition).}
\label{tab:constraints-a-inferred-charged-leptons}
\end{table}

\begin{table}
\centering
\begin{tabular}{|ccccccc|}
\toprule
$\mu\nu$                   & $\mathrm{Re}(c^{\ell})_{\mathrm{e}\upmu}^{\mu\nu}$ & $\mathrm{Re}(c^{\ell})_{\mathrm{e}\uptau}^{\mu\nu}$ & $\mathrm{Re}(c^{\ell})_{\upmu\uptau}^{\mu\nu}$ & $\mathrm{Im}(c^{\ell})_{\mathrm{e}\upmu}^{\mu\nu}$ & $\mathrm{Im}(c^{\ell})_{\mathrm{e}\uptau}^{\mu\nu}$ & $\mathrm{Im}(c^{\ell})_{\upmu\uptau}^{\mu\nu}$ \\
\midrule
$TT$                       & $5 \times 10^{-20}$                                & $5 \times 10^{-18}$                                 & $3 \times 10^{-27\,\diamond\dagger}$           & $5 \times  10^{-20\,\dagger}$                      & $5 \times 10^{-18}$                                 & $3 \times 10^{-27\,\diamond\dagger}$ \\
$TX$                       & $5 \times 10^{-23}$                                & $5 \times 10^{-18}$                                 & $5 \times 10^{-28}$                            & $5 \times 10^{-23}$                                & $5 \times 10^{-18}$                                 & $5 \times 10^{-23}$ \\
$TY$                       & $5 \times 10^{-23}$                                & $5 \times 10^{-18}$                                 & $5 \times 10^{-28}$                            & $5 \times 10^{-23}$                                & $5\times 10^{-18}$                                  & $5\times 10^{-23}$ \\
$TZ$                       & $5 \times 10^{-21}$                                & $5 \times 10^{-17}$                                 & ---                                            & $4\times 10^{-20\,\dagger}$ & $5\times 10^{-17}$   & --- \\
$XY$                       & $5 \times 10^{-22}$                                & $5 \times 10^{-18}$                                 & $5 \times 10^{-24}$                            & $5\times 10^{-22}$                                 & $5\times 10^{-18}$                                  & $5\times 10^{-22}$ \\
$XZ$                       & $5 \times 10^{-22}$                                & $5 \times 10^{-18}$                                 & $5 \times 10^{-24}$                            & $5\times 10^{-22}$                                 & $5\times 10^{-18}$                                  & $5\times 10^{-22}$ \\
$YZ$                       & $5 \times 10^{-22}$                                & $5 \times 10^{-17}$                                 & $5 \times 10^{-24}$                            & $5\times 10^{-22}$                                 & $5\times 10^{-17}$                                  & $5\times 10^{-22}$ \\
$XX$                       & $5 \times 10^{-22}$                                & $5 \times 10^{-17}$                                 & $5 \times 10^{-24}$                            & $5 \times  10^{-22}$                               & $5 \times 10^{-17}$                                 & $5 \times 10^{-22}$ \\
$YY$                       & $5 \times 10^{-22}$                                & $5 \times 10^{-17}$                                 & $5 \times 10^{-24}$                            & $5\times 10^{-22}$                                 & $5\times 10^{-17}$                                  & $5\times 10^{-22}$ \\
$ZZ$                       & $5 \times 10^{-20}$                                & $5 \times 10^{-17}$                                 & ---                                            & $2\times 10^{-19\,\dagger}$                        & $5\times 10^{-17}$                                  & --- \\
\bottomrule
\end{tabular}
\caption{The same as table~\protect\ref{tab:constraints-a-inferred-charged-leptons}, but for the dim-4 two-tensor coefficients $(c^{\ell})_{AB}^{\mu\nu}$ for charged leptons. We assumed that the right-handed charged-lepton fields remain unmodified: $(c_R)^{\mu\nu}_{AB}=0$. The unstated sensitivities to $(d^{\ell})_{AB}^{\mu\nu}$ equal those to $(c^{\ell})_{AB}^{\mu\nu}$. For the LR-symmetric scenario, $(c^{\ell})_{AB}^{\mu\nu}$ must be multiplied by 2, whereas no bounds can be inferred on $(d^{\ell})_{AB}^{\mu\nu}$. The symbol~$\dagger$ indicates data used from table~D30 of ref.~\cite{Kostelecky:2008ts}.}
\label{tab:constraints-c-inferred-charged-leptons}
\end{table}

Let us start with the minimal SME with the dim-3 coefficients $(a_{L,R})^{\mu}_{AB}$, $(a^{\ell})^{\mu}_{AB}$, $(b^{\ell})^{\mu}_{AB}$ as well as the dim-4 coefficients $(c_{L,R})^{\mu\nu}_{AB}$, $(c^{\ell})^{\mu\nu}_{AB}$, $(d^{\ell})^{\mu\nu}_{AB}$. Left-handed modifications in the neutrino sector that are constrained\footnote{Although these bounds originate from neutrino oscillations, in the spirit of ref.~\cite{Kostelecky:2011gq}, no right-handed sterile neutrinos are considered. Neutrino masses can be generated without introducing additional fields by adding a Weinberg operator \cite{Weinberg:1979sa} to eq.~(\ref{eq:lagrangian-sme-lepton-sector}).} from the absence of LV signals in neutrino oscillations (see table~\ref{tab:original-neutrino-constraints} of appendix~\ref{sec:derivation-constraints-details}) imply potential sensitivities for LV in the flavour-off-diagonal charged-lepton sector that are given in tables~\ref{tab:constraints-a-inferred-charged-leptons}, \ref{tab:constraints-c-inferred-charged-leptons}. Note that in LR-symmetric scenarios, the values for $(a^{\ell})_{AB}^{\mu}$ and $(c^{\ell})_{AB}^{\mu\nu}$ have to be multiplied by a factor of 2 while no sensitivities for $(b^{\ell})_{AB}^{\mu}$ and $(d^{\ell})_{AB}^{\mu\nu}$ can be obtained. Constraints on the coefficients in tables~\ref{tab:constraints-a-inferred-charged-leptons}, \ref{tab:constraints-c-inferred-charged-leptons} could have been determined previously directly only from processes that exhibit charged-lepton flavour violation. However, these bounds would supposedly be rather weak, as associated decay rates are expected to be suppressed by the square of the LV coefficient considered.

In an analog manner, limits on LV in the charged-lepton sector imply estimated sensitivities for neutrinos. It is known that the dim-3 coefficients $(a^{\ell})^{\mu}_{AA}$ in the absence of gravity are unobservable in experiments that involve a single lepton flavour only, as they can be removed by a field redefinition~\cite{Colladay:1996iz,Colladay:1998fq}. Therefore, there are no constraints on $(a^{\ell})^{\mu}_{AA}$ (see, e.g., tables~S2, D6 in ref.~\cite{Kostelecky:2008ts} for electrons).\footnote{The six diagonal $a^{\ell}$ and $a_L$ coefficients cannot be removed simultaneously. A field
redefinition acts on spinor space in the same way for all $a$-type coefficients present. This means that removing an $a$ coefficient for a certain flavour, say the electron, will produce this coefficient (with the opposite sign) for the remaining flavours (the muon and tau in this case). Therefore, it is the flavour-universal part that can be removed~\cite{Colladay:1996iz}, but differences like $(a^{\ell})^{\mu}_{\mathrm{ee}}-(a^{\ell})^{\mu}_{\upmu\upmu}$ could be constrained (which, to our knowledge, has not been done, yet).} So we must resort to the dim-3 coefficients $(b^{\ell})^{\mu}_{AA}$ (see table~\ref{tab:original-charged-lepton-constraints} in appendix~\ref{sec:derivation-constraints-details}) to deduce potential sensitivities on $(a_L)^{\mu}_{AA}$, which are stated in the first part of table~\ref{tab:constraints-inferred-neutrinos}.
\begin{table}
\centering
\begin{tabular}{|ccccc|}
\toprule
$d=3$ & $\mu$               & $(a_L)^{\mu}_{\mathrm{ee}}$     & $(a_L)^{\mu}_{\upmu\upmu}$                 & $(a_L)^{\mu}_{\uptau\uptau}$ \\
\midrule
      & $T$                 & $\unit[2 \times 10^{-27}]{GeV}$ & $\unit[2 \times 10^{-7}]{GeV}^{\dagger}$   & $\unit[2 \times 10^{-10}]{GeV^{\dagger}}$ \\
      & $X$                 & $\unit[2 \times 10^{-31}]{GeV}$ & $\unit[1 \times 10^{-25}]{GeV}^{\dagger}$  & $\unit[2 \times 10^{-10}]{GeV^{\dagger}}$ \\
      & $Y$                 & $\unit[2 \times 10^{-31}]{GeV}$ & $\unit[1 \times 10^{-25}]{GeV}^{\dagger}$  & $\unit[2 \times 10^{-10}]{GeV^{\dagger}}$ \\
      & $Z$                 & $\unit[2 \times 10^{-29}]{GeV}$ & $\unit[4 \times 10^{-23}]{GeV}^{\dagger}$  & $\unit[2 \times 10^{-10}]{GeV^{\dagger}}$ \\
\midrule
$d=4$ & $\mu\nu$            & $(c_L)^{\mu\nu}_{\mathrm{ee}}$  &  $(c_L)^{\mu\nu}_{\upmu\upmu}$             & $(c_L)^{\mu\nu}_{\uptau\uptau}$ \\
\midrule
      & $TT$                & ---                             & ---                                        & ---                          \\
      & $TX$                & $1 \times 10^{-15}$             & $1\times 10^{-11\,\dagger}$                & ---                          \\
      & $TY$                & $1 \times 10^{-15}$             & $1\times 10^{-11\,\dagger}$                & ---                          \\
      & $TZ$                & $1 \times 10^{-17}$             & $1\times 10^{-11\,\dagger}$                & ---                          \\
      & $XY$                & $1 \times 10^{-17}$             & ---                                        & ---                          \\
      & $XZ$                & $1 \times 10^{-18}$             & ---                                        & ---                          \\
      & $YZ$                & $1 \times 10^{-18}$             & ---                                        & ---                          \\
      & $XX$                & $3 \times 10^{-15}$             & ---                                        & ---                          \\
      & $YY$                & $3 \times 10^{-15}$             & ---                                        & ---                          \\
      & $ZZ$                & $7 \times 10^{-15}$             & ---                                        & ---                          \\
\bottomrule
\end{tabular}
\caption{Two-sided potential sensitivities on flavour-diagonal coefficients in the neutrino sector obtained from the charged-lepton sector via $\mathit{SU}(2)_L$ gauge invariance (see appendix~\ref{sec:derivation-constraints-details} for calculational details). For the values for $(a_L)^{\mu}_{AA}$ we assumed that only left-handed fields are modified such that $(a_R)^{\mu}_{AA}=0$. Note that the LR-symmetric scenario does not provide values for $(a_L)^{\mu}_{AA}$ while those on $(c_L)^{\mu}_{AA}$ are independent of any assumption on the modification of right-handed charged leptons.}
\label{tab:constraints-inferred-neutrinos}
\end{table}

However, both dim-4 coefficients $(c^{\ell})^{\mu\nu}_{AA}$ and $(d^{\ell})^{\mu\nu}_{AA}$ are physical,\footnote{Local gauge invariance couples eq.~(\ref{eq:lagrangian-sme-lepton-sector}) to photons. Note that the flavour-universal part of the $c$ coefficients can be mapped onto the dim-4 {\em CPT}-even nonbirefringent photon coefficients and vice versa~\cite{Bailey:2004na,Altschul:2006zz}. However, flavour-nonuniversal effects cannot be transformed to other sectors.} and strict bounds on them are given in ref.~\cite{Kostelecky:2008ts}. Taking advantage of this, we express $(c_R)^{\mu\nu}_{AA}$, $(c_L)^{\mu\nu}_{AA}$ in terms of the charged-lepton coefficients:
\begin{subequations}
\label{eq:cR-cL-from-cell-dell}
\begin{align}
(c_R)^{\mu\nu}_{AA}&=(c^{\ell})^{\mu\nu}_{AA}-(d^{\ell})^{\mu\nu}_{AA}\,, \displaybreak[0]\\[2ex]
(c_L)^{\mu\nu}_{AA}&=(c^{\ell})^{\mu\nu}_{AA}+(d^{\ell})^{\mu\nu}_{AA}\,,
\end{align}
\end{subequations}
which allows us to constrain both $(c_L)^{\mu\nu}_{AA}$ and $(c_R)^{\mu\nu}_{AA}$. Note that only $(c_L)^{\mu\nu}_{AA}$ is related to LV in the neutrino sector, whereas $(c_R)^{\mu\nu}_{AA}$ modifies right-handed charged leptons. The computed values are provided in the second part of table~\ref{tab:constraints-inferred-neutrinos} and are based on the numbers listed in table~\ref{tab:original-charged-lepton-constraints} of appendix~\ref{sec:derivation-constraints-details}. Some of these limits are derived from bounds on linear combinations of coefficients, stated in ref.~\cite{Kostelecky:2008ts} (see the definitions in tables~P48, P49 of the latter reference). In these cases, all coefficients are set to zero except for those that we are interested in. This procedure is widely accepted, as it prevents unnatural cancelations between different types of LV. Whenever possible, we used the relations from table~P49 of ref.~\cite{Kostelecky:2008ts} that would allow us to derive the sensitivities by setting as few coefficients to zero as possible. These procedures imply that the obtained numbers must be interpreted as estimated experimental sensitivities instead of strict constraints on LV.

Besides, some additional assumptions are taken. First, we assume $c^{\mu\nu}$, $d^{\mu\nu}$ to be symmetric in the Lorentz indices, as effects related to antisymmetric combinations are usually suppressed (see, e.g., the modified dispersion relation for the $c$ coefficients given in ref.~\cite{Altschul:2006uw} or classical-particle descriptions of LV at leading order in refs.~\cite{Reis:2017ayl,Edwards:2018lsn,Schreck:2019mmr}). Furthermore, we assume $c^{TT}=0$ in the muon and tau sector for the bounds stated in ref.~\cite{Altschul:2006uw} because of the following reason. The parameter space for a symmetric and traceless $c_{\mu\nu}$ is eight-dimensional. An inequality for a combination of these coefficients rules out one half of the parameter space separated by a seven-dimensional hyperplane. A sufficient number of high-energy cosmic-ray or photon events coming from different directions can constrain the coefficients within a bounded polytope in the parameter space. Negative values for $c^{TT}$ cannot be constrained by considering just one particular type of exotic decay process (e.g., photon decay). By taking $c^{TT}=0$, all one-sided bounds stated in ref.~\cite{Altschul:2006uw} are rendered two-sided (see ref.~\cite{Altschul:2008qg}, in addition). Similar arguments were also employed in other analyses such as in ref.~\cite{Diaz:2013wia}.

The sensitivities for $(a_L)^{\mu}_{AA}$ are valid when LV only affects left-handed charged leptons while in the LR-symmetric case no sensitivities can be inferred for them. It is an advantageous property of the $c$ coefficients that sensitivities for $(c_L)^{\mu\nu}_{AA}$ can be obtained without any additional assumptions. Therefore, we resort to the available charged-lepton sensitivities in the second part of table~\ref{tab:constraints-inferred-neutrinos}, which is why several undetermined entries remain, in particular, in the tau sector. In these cases there is not enough information available on $(d^{\ell})_{AA}^{\mu\nu}$ that allows us to compute $(c_L)^{\mu\nu}_{AA}$. This observation poses a great motivation for experimentalists to conceive experiments that look for $d$-type LV in the muon and tau sector.

Moreover, the stringent bounds from flavour-diagonal charged leptons originate, e.g., from high-precision experiments with electrons (such as Penning traps and spectroscopy; see ref.~\cite{Kostelecky:2008ts}). Importantly, the inferred sensitivities for LV in neutrinos exceed some of the known constraints by several orders of magnitude (cf. the constraints on the flavour-universal, isotropic coefficients $\ring{a}^{(3)}$ and $\ring{c}^{(4)}$ in table~S4 of ref.~\cite{Kostelecky:2008ts}), i.e., sensitivity is gained in the flavour-diagonal neutrino sector. In this case, sensitivity for LV for right-handed charged leptons approximately corresponds to that valid for neutrinos, $(c_R)^{\mu\nu}_{AA}\approx (c_L)^{\mu\nu}_{AA}$, showing that parity violation for this particular $c$-type background field is highly suppressed.

A final caveat has to be given at this point. The minimal SME involves Yukawa-like couplings between the Higgs and the lepton fields (see, e.g., the vector-valued coefficients $I^{\mu}$ and the pseudovector-valued coefficients $J^{\mu}$ in eqs.~(34), (37) of ref.~\cite{Ferrero:2011yu}). Spontaneous symmetry breaking can provide dim-3 operators in the low-energy limit that are proportional to the vacuum expectation value $v$ of the Higgs field. These operators violate $\mathit{SU}(2)_L$ invariance and throughout the manuscript it is assumed that Yukawa-like couplings of this type are suppressed in comparison to the controlling coefficients of the free lepton sector.

\section{Consequences for time-of-flight neutrino analysis}
\label{sec:consequences-neutrino-analyses}

Neutrinos allow for precise tests of Lorentz invariance. Since they interact very weakly and travel long distances before interacting, they are sensitive to any kind of background that modifies their propagation properties. In fact, a broad series of searches for LV in the neutrino sector has been carried out, see, e.g.,~\cite{Katori:2006mz,Diaz:2009qk,Katori:2010nf,Diaz:2010ft,AguilarArevalo:2011yi,Diaz:2011ia,Katori:2011zz,Abe:2012gw,Katori:2012pe,Diaz:2013saa,Diaz:2013wia,Diaz:2013ywa,Diaz:2014yva,Diaz:2014hca,Diaz:2015dxa,Diaz:2015aua,Diaz:2016fqd,Diaz:2016xpw,Arguelles:2016rkg,Katori:2016eni,Abe:2017eot,Aartsen:2017ibm}. Over the last years the IceCube experiment~\cite{Ahrens:2003ix} has detected a collection of neutrinos with energies in the TeV and even PeV regime~\cite{Aartsen:2013bka,Aartsen:2013jdh,Aartsen:2014gkd,Aartsen:2016ngq}, allowing for tests of Lorentz symmetry in previously uncharted regions. A subset of the IceCube neutrino events is assumed to originate from gamma ray bursts (GRBs)~\cite{Eichler:1989ve,Paczynski:1994uv,Waxman:1995vg}, opening up the possibility of testing LV by comparing photon and neutrino arrival times~\cite{Jacob:2006gn,Jacob:2008bw}. In fact, statistically significant hints for in-\textit{vacuo} modified dispersion relations for GRB neutrinos~\cite{Amelino-Camelia:2016fuh,Amelino-Camelia:2016ohi,Amelino-Camelia:2017zva,Huang:2018ham,Huang:2019etr} as well as GRB photons~\cite{Zhang:2014wpb,Xu:2016zxi,Xu:2016zsa} have been exposed. However, as LV in the photon sector is too tightly constrained (see ref.~\cite{Kostelecky:2008ts}), it is usually not considered to explain arrival time differences between different particle species.

In refs.~\cite{Huang:2018ham,Huang:2019etr} it was shown that a modified dispersion relation for neutrinos of the form
\begin{equation}
\label{eq:modified-dispersion-relation}
E=E_0\left[1\pm\frac{1}{2}\left(\frac{E_0}{E_{\mathrm{LV}}}\right)\right]\,,
\end{equation}
with $E_0=|\vec p|\equiv p$, where $\vec{p}$ is the spatial neutrino momentum, accounts for the data exceptionally well. Here, $E_{\mathrm{LV}}$ is the characteristic energy scale associated with the fundamental physics that may induce LV. The upper sign can refer to neutrinos and the lower one to antineutrinos (or vice versa), meaning that neutrinos are superluminal, whereas antineutrinos are subliminal. Since IceCube cannot distinguish between neutrinos and antineutrinos, it is unknown which sign holds for particles and which one for antiparticles. Superluminal neutrinos\footnote{Note that superluminal particles are not necessarily in conflict with microcausality in theories with broken Lorentz invariance as shown in ref.~\cite{Kostelecky:2000mm} for Dirac fermions and in refs.~\cite{Klinkhamer:2010zs,Schreck:2011ai,Schreck:2013gma} for photons. Microcausality is valid as long as signals do not propagate outside of modified mass shells or light cones.}~\cite{Alexandre:2011bu,Giudice:2011mm,Dvali:2011mn,AmelinoCamelia:2011dx,Cacciapaglia:2011ax,Bi:2011nd} were already considered in the context of the former OPERA anomaly~\cite{Adam:2011faa}.

The coefficients that lead to a modified neutrino dispersion relation with the same energy-dependence as that given in eq.~\ref{eq:modified-dispersion-relation}, are the dim-5 $a$ coefficients. Since flavour off-diagonal ones are strongly constrained by neutrino oscillations~\cite{Aartsen:2017ibm}, we assume them to be flavour-universal: $(a^{(5)}_L)^{\alpha\kappa\lambda}_{AB}\equiv (a_L^{(5)})^{\alpha\kappa\lambda}\delta_{AB}$.
Furthermore, considering a coordinate frame where LV is isotropic, $a_L^{(5)000}\equiv \ring{a}^{(5)}_{L,0}$ and $a_L^{(5)0jj}=a_L^{(5)j0j}=a_L^{(5)jj0}\equiv \ring{a}^{(5)}_{L,2}/3$ for $j=1,2,3$, the modified neutrino dispersion relation at first order in LV reads
\begin{equation}
\label{eq:modified-dispersion-relation-sme}
E\simeq p\left[1+(\ring{a}_{L,0}^{(5)}+\ring{a}_{L,2}^{(5)})p\right]\,,
\end{equation}
and we find the correspondence
\begin{equation}
\pm\frac{1}{2E_{\mathrm{LV}}}=\ring{a}^{(5)}_{L,0}+\ring{a}^{(5)}_{L,2}\,.
\end{equation}
In ref.~\cite{Huang:2018ham} the value $E_{\mathrm{LV}}=\unit[6.5\times 10^{17}]{GeV}$ was obtained as the energy scale where LV effects are generated. The procedure was to fit a straight line through a sample of data points which related the arrival time differences between GRB photons and high-energy cosmic
neutrinos with the neutrino energies. Details on this fit are presented in appendix~\ref{sec:fit-neutrino-time-of-flight}. The energy scale derived in this way translates into the following values for the combination of isotropic dim-5 $a$ coefficients:
\begin{equation}
\label{eq:values-isotropic-coefficients}
\ring{a}^{(5)}_{L,0}+\ring{a}^{(5)}_{L,2}=\pm\unit[7.7\times 10^{-19}]{GeV^{-1}}\,.
\end{equation}
The latter numbers were also obtained in ref.~\cite{Zhang:2018otj}. Flavour-universal LV described by an isotropic coefficient $\ring{a}^{(5)}$ is constrained at the level of $\unit[10^{-18}]{GeV^{-1}}$ (cf.~table~S4 in ~\cite{Kostelecky:2008ts}). Thus, the value of eq.~(\ref{eq:values-isotropic-coefficients}) is not in conflict with existing constraints in the neutrino sector.

Restricting the dim-5 operator $(\hat{a}_L)^{\mu}_{AB}$ to its flavour-universal and isotropic parts, we obtain the following dim-5 coefficients in the charged-lepton sector:%
\begin{equation}
\label{eq:dim-5-coefficients-charged-lepton}
\ring{a}^{(5)\ell}_{AB}=\frac{1}{2}\left[\ring{a}^{(5)}_{L,0}+\ring{a}^{(5)}_{L,2}\right]\delta_{AB}=\ring{b}^{(5)\ell}_{AB}\,.
\end{equation}
The coefficient $\ring{a}^{(5)\ell}_{ee}$ is tightly bounded in the ultra-relativistic limit (cf.~ref.~\cite{Kostelecky:2013rta} and table~D7 in~ref.~\cite{Kostelecky:2008ts}). There is a weaker two-sided constraint ranging from around $\unit[10^{-20}]{GeV^{-1}}$ to $\unit[3\times 10^{-17}]{GeV^{-1}}$~\cite{Altschul:2010nf}. The latter is a consequence of the absence of photon decay for $\unit[50]{TeV}$ photons originating from the Crab Nebula as well as the absence of vacuum Cherenkov radiation losses at LEP. Much better constraints at the level of $\unit[10^{-25}]{GeV^{-1}}$ were obtained via the absence of LV signals in the broad-band spectrum of the Crab Nebula~\cite{Maccione:2007yc}. Detailed studies of the synchrotron radiation spectrum of the Crab Nebula even improved these bounds by another two orders of magnitude~\cite{Konopka:2002tt,Jacobson:2002ye}. Last but not least, remarkable bounds at the level of $\unit[10^{-34}]{GeV^{-1}}$ were found by the absence of energy losses of ultra-high-energy electrons that might be caused by LV~\cite{Gagnon:2004xh}. This is why eq.~(\ref{eq:values-isotropic-coefficients}) clashes with existing limits for charged leptons.

A second argument showing that eq.~(\ref{eq:values-isotropic-coefficients}) is in conflict with the detection of PeV neutrinos was developed in ref.~\cite{Zhang:2018otj}. Superluminal neutrinos lose energy by emission of electron-positron pairs via an intermediate Z boson if their energy is above a certain threshold. If LV had the size quoted in eq.~(\ref{eq:values-isotropic-coefficients}) for PeV neutrinos, the latter would have lost a major part of their energy before being detected on Earth.\footnote{More details on such radiation processes can be found in refs.~\cite{Gagnon:2004xh,Cohen:2011hx,Bezrukov:2011qn,Kostelecky:2011gq,Diaz:2013wia,Somogyi:2019yis}.} However, eq.~(\ref{eq:dim-5-coefficients-charged-lepton}) applies to both superluminal and subluminal neutrinos and therefore rules out an explanation of the IceCube time lag via the dim-5 operator $(\hat{a}_L)^{\mu}_{AB}$.

\section{Conclusions and outlook}
\label{sec:conclusions}

In this paper we showed that $\mathit{SU}(2)_L$ gauge invariance of the SME allows us to estimate sensitivities for LV in the flavour-diagonal neutrino sector from existing ones in the charged-lepton sector. The inferred sensitivities are superior to those determined by experiment. Furthermore, flavour-changing modifications of the charged-lepton sector, previously unconstrained, can be bounded via the neutrino sector. Due to a lack of bounds on $d$ coefficients for muons and taus, a larger number of coefficients $(c_L)^{\mu\nu}_{\upmu\upmu}$, $(c_L)^{\mu\nu}_{\uptau\uptau}$ remains unconstrained. This finding provides motivation for determining experimental limits on the muon and tau $d$ coefficients such that compilations like table~\ref{tab:constraints-inferred-neutrinos} can be complemented in the future.

As a particular application, we used this method to assess the validity of an explanation by LV of the arrival time difference between photons from GRBs and correlated neutrinos detected by IceCube. While in this context superluminal neutrinos were already excluded by electron-positron radiation via Z effects, we show that subluminal modifications of the neutrino velocity are also in conflict with existing bounds. An analogous argument could have ruled out a wide range of explanations for the OPERA anomaly~\cite{Adam:2011faa}, which was at the level of $10^{-5}$.

However, note that our bounds could be avoided by using higher-dimensional operators, e.g., a generalized Weinberg operator \cite{Weinberg:1979sa}:
\begin{equation}
\label{eq:modified-weinberg-operator}
{\cal L}_{}^{(7)}=-a_{L,AB}^{(7)\mu\alpha\beta}\phi _I^*{\phi _J}{\varepsilon_{II'}}{\varepsilon_{JJ'}}\bar L_A^{I'}(\mathrm{i}{\partial_\alpha})(\mathrm{i}{\partial_\beta}){\gamma_\mu}L_B^{J'}\,,
\end{equation}
with the SM Higgs doublet $\phi$, the totally antisymmetric Levi-Civita symbol $\varepsilon_{IJ}$, and $\mathit{SU}(2)_L$ indices $I^{(\prime)}$, $J^{(\prime)}$. Here, LV originates from a {\em CPT}-violating interaction. However, modified dispersion relations of neutrinos are only induced when the Higgs field acquires its vacuum expectation value. Interestingly, a modification of the Weinberg operator similar to that of eq.~(\ref{eq:modified-weinberg-operator}) was considered in ref.~\cite{Klinkhamer:2012uu} with a two-tensor-valued background field. The latter can be generated via a composite operator formed from two gradients of an additional scalar isosinglet introduced into the SM. An analogous mechanism employing three gradients of this scalar field is expected to generate the background field giving rise to a nonzero $a_{L,AB}^{(7)\mu\alpha\beta}$.

Similar arguments like those put forward in this article resting on $\mathit{SU}(2)_L$ gauge invariance link constraints between left-handed up and down-type quarks. Here, one can expect to constrain modifications to up and charm quarks from Kaon mixing as well as the largely unconstrained top-quark sector (see ref.~\cite{Carle:2019ouy} for a recent account on bounds from LHC searches) from LV bounds from $\mathrm{B_s}$-$\mathrm{\bar B_s}$ and $\mathrm{B_d}$-$\mathrm{\bar B_d}$ mixing~\cite{Kostelecky:1994rn,Edwards:2019lfb} opening up interesting future lines of research.

\medskip

\acknowledgments

The authors are grateful to V.A.~Kosteleck\'{y} and M.~Spira for valuable comments on the manuscript. Furthermore, they thank B.-Q.~Ma for helpful discussions as to refs.~\cite{Huang:2018ham,Huang:2019etr}, N.~Russell for clarifying certain aspects of the data tables~\cite{Kostelecky:2008ts} as well as B.~Altschul for useful remarks on the constraints in ref.~\cite{Altschul:2006uw}. The work of A.C. is supported by a Professorship Grant (PP00P2\_176884) of the Swiss National Science Foundation. M.S. is indebted to FAPEMA Universal 01149/17, FAPEMA Universal 00830/19, CNPq Universal 421566/2016-7, CNPq Produtividade 312201/2018-4, and CAPES/Finance Code 001 for support.

\appendix

\section{Details on deriving the sensitivities}
\label{sec:derivation-constraints-details}

We take the opportunity of stating further calculational details that are not of direct importance for understanding the implications of our results stated in tables~\ref{tab:constraints-a-inferred-charged-leptons}, \ref{tab:constraints-c-inferred-charged-leptons}, and \ref{tab:constraints-inferred-neutrinos} of section~\ref{sec:connection-coefficients}. Under the assumption of $(\hat{a}_R)^{\mu}_{AB}=(\hat{c}_R)^{\mu\nu}_{AB}=0$, the following connections arise between operators in the neutrino and the charged-lepton sector:
\begin{subequations}
\label{eq:connection-constraints-1}
\begin{align}
(\hat{a}^{\ell})^{\mu}_{AB}&=\frac{1}{2}(\hat{a}_L)^{\mu}_{AB}=(\hat{b}^{\ell})^{\mu}_{AB}\,, \displaybreak[0]\\[2ex]
(\hat{c}^{\ell})^{\mu\nu}_{AB}&=\frac{1}{2}(\hat{c}_L)^{\mu\nu}_{AB}=(\hat{d}^{\ell})^{\mu\nu}_{AB}\,.
\end{align}
\end{subequations}
Inverting the latter provides additional relations:
\begin{subequations}
\label{eq:connection-constraints-2}
\begin{align}
\label{eq:connection-constraints-aL-aell}
(\hat{a}_L)^{\mu}_{AB}&=2(\hat{a}^{\ell})^{\mu}_{AB}\,, \displaybreak[0]\\[2ex]
\label{eq:connection-constraints-aL-bell}
(\hat{a}_L)^{\mu}_{AB}&=2(\hat{b}^{\ell})^{\mu}_{AB}\,, \displaybreak[0]\\[2ex]
\label{eq:connection-constraints-cL-cell}
(\hat{c}_L)^{\mu\nu}_{AB}&=2(\hat{c}^{\ell})^{\mu\nu}_{AB}\,, \displaybreak[0]\\[2ex]
\label{eq:connection-constraints-cL-dell}
(\hat{c}_L)^{\mu\nu}_{AB}&=2(\hat{d}^{\ell})^{\mu\nu}_{AB}\,.
\end{align}
\end{subequations}
While working in the minimal SME, let there be a certain set of two-sided constraints for the vector and pseudovector coefficients in the neutrino and charged-lepton sector:
\begin{subequations}
\label{eq:two-sided-constraints-a}
\begin{align}
-\widetilde{X}^{\mu}_{AB}&<(a_L)^{\mu}_{AB}<X^{\mu}_{AB}\,, \displaybreak[0]\\[2ex]
\label{eq:two-sided-constraints-a-Y}
-\widetilde{Y}^{\mu}_{AB}&<(a^\ell)^{\mu}_{AB}<Y^{\mu}_{AB}\,, \displaybreak[0]\\[2ex]
\label{eq:two-sided-constraints-a-Z}
-\widetilde{Z}^{\mu}_{AB}&<(b^{\ell})^{\mu}_{AB}<Z^{\mu}_{AB}\,.
\end{align}
\end{subequations}
An analogous set of two-sided constraints shall exist for the two-tensor coefficients:
\begin{subequations}
\label{eq:two-sided-constraints-c}
\begin{align}
-\widetilde{X}^{\mu\nu}_{AB}&<(c_L)^{\mu\nu}_{AB}<X^{\mu\nu}_{AB}\,, \displaybreak[0]\\[2ex]
\label{eq:two-sided-constraints-c-Y}
-\widetilde{Y}^{\mu\nu}_{AB}&<(c^\ell)^{\mu\nu}_{AB}<Y^{\mu\nu}_{AB}\,, \displaybreak[0]\\[2ex]
\label{eq:two-sided-constraints-c-Z}
-\widetilde{Z}^{\mu\nu}_{AB}&<(d^{\ell})^{\mu\nu}_{AB}<Z^{\mu\nu}_{AB}\,.
\end{align}
\end{subequations}
\begin{table}[t]
\centering
	\begin{tabular}{|cccccc|}
		\toprule
		New constraint & \multicolumn{3}{c}{Inferred from} & Table & Ref. \\
		\midrule
		$\mathrm{Im}(a^{\ell})_{\mathrm{e}\upmu}^{T}$  & $|\mathrm{Im}(a_L)_{\mathrm{e}\upmu}^{T}|$    & $<$ & $\unit[4.2\times 10^{-20}]{GeV}$ & D29(2) & \cite{Katori:2012pe} \\
		$\mathrm{Im}(a^{\ell})_{\upmu\uptau}^{T}$      & $\mathrm{Im}(a_L)_{\upmu\uptau}^{T}$          & $<$ & $\unit[5.1\times 10^{-24}]{GeV}$ & D29(4) & \cite{Abe:2014wla} \\
		$\mathrm{Im}(c^{\ell})_{\mathrm{e}\upmu}^{TT}$ & $|\mathrm{Im}(c_L)_{\mathrm{e}\upmu}^{TT}|$   & $<$ & $9.6\times 10^{-20}$             & D30(4) & \cite{Katori:2012pe} \\
		$\mathrm{Re}(c^{\ell})_{\upmu\uptau}^{TT}$     & $\mathrm{Re}(c_L)_{\upmu\uptau}^{TT}$         & $<$ & $5.8\times 10^{-27}$             & D30(6) & \cite{Abe:2014wla} \\
		$\mathrm{Im}(c^{\ell})_{\upmu\uptau}^{TT}$     & $\mathrm{Im}(c_L)_{\upmu\uptau}^{TT}$         & $<$ & $5.6\times 10^{-27}$             & D30(6) & \cite{Abe:2014wla} \\
        $\mathrm{Im}(c^{\ell})_{\mathrm{e}\upmu}^{TZ}$ & $|\mathrm{Im}(c_L)_{\mathrm{e}\upmu}^{TZ}|$   & $<$ & $7.8\times 10^{-20}$             & D30(4) & \cite{Katori:2012pe} \\
        $\mathrm{Im}(c^{\ell})_{\mathrm{e}\upmu}^{ZZ}$ & $|\mathrm{Im}(c_L)_{\mathrm{e}\upmu}^{ZZ}|$   & $<$ & $3.4\times 10^{-19}$             & D30(4) & \cite{Katori:2012pe} \\
		\bottomrule
	\end{tabular}
\caption{Coefficients on which new constraints can be inferred (first column), coefficients in the neutrino sector from which the constraints were inferred (second column) along with the particular tables of ref.~\cite{Kostelecky:2008ts} that the values were taken from (third column), and the original references (fourth column). Only the bounds that are not stated in the summary tables of ref.~\cite{Kostelecky:2008ts} are given. Pure laboratory experiments were prioritised over experiments involving cosmic neutrinos.}
\label{tab:original-neutrino-constraints}
\end{table}
\begin{table}
\centering
	\begin{tabular}{|lcccccl|}
		\toprule
		Inferred                                                  & \multicolumn{3}{c}{Experimental limit}     & Original coefficients                                      & Table & Ref.
		\\
		\midrule
		$(a_L)_{\mathrm{ee}}^T$            & $|b^T_{\mathrm{ee}}|$ & $<$ & $\unit[10^{-27}]{GeV}$              & $\tilde{g}_T$                                              & S2 & \cite{Kostelecky:2008ts} \\
		$(a_L)_{\mathrm{ee}}^{X,Y}$        & $|b^{X,Y}_{\mathrm{ee}}|$ & $<$ & $\unit[10^{-31}]{GeV}$          & $\tilde{b}_X$, $\tilde{b}_Y$                               & S2 & \cite{Kostelecky:2008ts} \\
        $(a_L)_{\mathrm{ee}}^Z$            & $|b^Z_{\mathrm{ee}}|$ & $<$ & $\unit[10^{-29}]{GeV}$              & $\tilde{b}_Z$                                              & S2 & \cite{Kostelecky:2008ts} \\
		$(c_L)_{\mathrm{ee}}^{TT}$         & $|c^{TT}_{\mathrm{ee}}|$ & $<$ & $2.0\times 10^{-16}$             & $\tilde{c}_{TT}$                                           & S2 & \cite{Kostelecky:2008ts} \\
		$(c_L)_{\mathrm{ee}}^{TX,TY}$      & $|c^{TX,TY}_{\mathrm{ee}}|$ & $<$ & $9.8\times 10^{-16}$          & $\tilde{c}_{TX}$, $\tilde{c}_{TY}$                         & S2 & \cite{Kostelecky:2008ts} \\
		                                   & $|d^{TX,TY}_{\mathrm{ee}}|$ & $<$ & $2.0\times 10^{-28}$          & $\tilde{b}_X$, $\tilde{b}_Y$                               & S2 & \cite{Kostelecky:2008ts} \\
        $(c_L)_{\mathrm{ee}}^{TZ}$         & $|c^{TZ}_{\mathrm{ee}}|$ & $<$ & $9.8\times 10^{-18}$             & $\tilde{c}_{TZ}$                                           & S2 & \cite{Kostelecky:2008ts} \\
                                           & $|d_{\mathrm{e}}^{TZ}|$ & < & $2.0\times 10^{-26}$                & $\tilde{b}_Z$                                              & S2 & \cite{Kostelecky:2008ts} \\
        $(c_L)_{\mathrm{ee}}^{XY}$         & $|c^{XY}_{\mathrm{ee}}|$ & $<$ & $9.8\times 10^{-18}$             & $\tilde{c}_Z$                                              & S2 & \cite{Kostelecky:2008ts} \\
                                           & $|d^{XY}_{\mathrm{ee}}|$ & $<$ & $9.8\times 10^{-24}$             & $\tilde{d}_{XY}$                                           & S2 & \cite{Kostelecky:2008ts} \\
		$(c_L)_{\mathrm{ee}}^{XZ,YZ}$      & $|c^{XZ,YZ}_{\mathrm{ee}}|$ & $<$ & $9.8\times 10^{-19}$          & $\tilde{c}_Y$, $\tilde{c}_X$                               & S2 & \cite{Kostelecky:2008ts} \\
		                                   & $|d^{XZ,YZ}_{\mathrm{ee}}|$ & $<$ & $9.8\times 10^{-24}$          & $\tilde{d}_{XZ}$, $\tilde{d}_{YZ}$                         & S2 & \cite{Kostelecky:2008ts} \\
		$(c_L)_{\mathrm{ee}}^{XX,YY}$      & $|c^{XX,YY}_{\mathrm{ee}}|$ & $<$ & $3.3\times 10^{-15}$          & $\tilde{c}_Q$, $\tilde{c}_-$, $\tilde{c}_{TT}$             & S2 & \cite{Kostelecky:2008ts} \\
                                           & $|d^{XX,YY}_{\mathrm{ee}}|$ & $<$ & $1.1\times 10^{-23}$          & $\tilde{d}_+$, $\tilde{d}_-$                               & S2 & \cite{Kostelecky:2008ts} \\
        $(c_L)_{\mathrm{ee}}^{ZZ}$         & $|c_{\mathrm{ee}}^{ZZ}|$ & < & $6.6\times 10^{-15}$               & $\tilde{c}_Q$, $\tilde{c}_{TT}$                            & S2 & \cite{Kostelecky:2008ts} \\
                                           & $|d_{\mathrm{ee}}^{ZZ}|$ & < & $2.2\times 10^{-23}$               & $\tilde{b}_T$, $\tilde{g}_T$, $\tilde{d}_+$, $\tilde{d}_Q$ & S2 & \cite{Kostelecky:2008ts} \\
		\midrule
		$(a_L)_{\upmu\upmu}^T$             & $|b^T_{\upmu\upmu}|$ & $<$ & $\unit[1.1\times 10^{-7}]{GeV}$      & $b^T$                                                      & D23 & \cite{Noordmans:2014hxa}* \\
		$(a_L)_{\upmu\upmu}^{X,Y}$         & $|b^{X,Y}_{\upmu\upmu}|$ & $<$ & $\unit[4.8\times 10^{-26}]{GeV}$ & $\check{b}_X^+$, $\check{b}_Y^+$                           & D23 & \cite{Bennett:2007aa} \\
        $(a_L)_{\upmu\upmu}^Z$             & $|b^Z_{\upmu\upmu}|$ & $<$ & $\unit[2.2\times 10^{-23}]{GeV}$     & $b_Z$                                                      & D23 & \cite{Bennett:2007aa} \\
		$(c_L)_{\upmu\upmu}^{TT}$          & $c^{TT}_{\upmu\upmu}$ & $=$ & 0                                   & assumption                                                 & --- & \cite{Altschul:2006uw}* \\
		$(c_L)_{\upmu\upmu}^{TJ}$          & $|c^{TJ}_{\upmu\upmu}|$ & $<$ & $10^{-11}$                        & $|c|$                                                      & D24 & \cite{Altschul:2006uw}* \\
		                                   & $|d^{TX,TY}_{\upmu\upmu}|$ & $<$ & $2\times 10^{-22}$             & $\tilde{b}_X$, $\tilde{b}_Y$                               & D23 & \cite{Hughes:2001yk} \\
		                                   & $|d^{TZ}_{\upmu\upmu}|$ & $<$ & $1.1\times 10^{-21}$              & $d_{Z0}$                                                   & D23 & \cite{Bennett:2007aa}\\
		$(c_L)_{\upmu\upmu}^{JK}$          & $|c^{JK}_{\upmu\upmu}|$ & $<$ & $10^{-11}$                        & $|c|$                                                      & D24 & \cite{Altschul:2006uw}* \\
		\midrule
		$(a_L)_{\uptau\uptau}^T$           & $|b^T_{\uptau\uptau}|$ & $<$ & $\unit[8.5\times 10^{-11}]{GeV}$   & $|b^{\mu}|$                                                & D26 & \cite{Escobar:2018hyo}* \\
		$(a_L)_{\uptau\uptau}^J$           & $|b^J_{\uptau\uptau}|$ & $<$ & $\unit[8.5\times 10^{-11}]{GeV}$   & $|b^{\mu}|$                                                & D26 & \cite{Escobar:2018hyo}* \\
		$(c_L)_{\uptau\uptau}^{TT}$        & $c^{TT}_{\uptau\uptau}$ & $=$ & 0                                 & assumption                                                 & --- & \cite{Altschul:2006uw}* \\
        $(c_L)_{\uptau\uptau}^{TJ}$        & $|c^{TJ}_{\uptau\uptau}|$ & $<$ & $10^{-8}$                       & $|c|$                                                      & D26 & \cite{Altschul:2006uw}* \\
	    $(c_L)_{\uptau\uptau}^{JK}$        & $|c^{JK}_{\uptau\uptau}|$ & $<$ & $10^{-8}$                       & $|c|$                                                      & D26 & \cite{Altschul:2006uw}* \\
		\bottomrule
	\end{tabular}
\caption{Neutrino coefficients, on which new constraints can be inferred (first column), coefficients in the charged-lepton sector from which the constraints were inferred (second column) along with the particular coefficients (third column) and tables (fourth column) of ref.~\cite{Kostelecky:2008ts} that the values were deduced from, and the original references (fifth column). Whenever possible, summary table entries were prioritised over data table entries and bounds from pure laboratory experiments were prioritised over astrophysical limits. Theory papers are indicated by an asterisk. Furthermore, $J$, $K$ are generic spacelike indices. To avoid unnatural cancelations between coefficients, we chose suitable combinations of lower and upper limits. For example, in $md_{ZZ}=(\tilde{b}_T-\tilde{g}_T+\tilde{d}_+-\tilde{d}_Q)/2$ we employed lower limits on $\tilde{g}_T$, $\tilde{d}_Q$ to obtain upper limits on $d_{ZZ}$. Otherwise, from the existing positive numbers in table~S2 of ref.~\cite{Kostelecky:2008ts} we would have immediately deduced that $d_{ZZ}=0$ for electrons.}
\label{tab:original-charged-lepton-constraints}
\end{table}
For simplicity, we will employ the same variables to denote the bounds for vector and two-tensor coefficients. The number of Lorentz indices allows for the distinction between them. Imposing eqs.~(\ref{eq:connection-constraints-1}), we can infer new constraints for charged leptons from the bounds on LV in the neutrino sector:
\begin{subequations}
\label{eq:inferred-bounds-a}
\begin{align}
-\frac{1}{2}\tilde{X}^{\mu}_{AB}&<(a^\ell)^{\mu}_{AB}<\frac{1}{2}X^{\mu}_{AB}\,, \displaybreak[0]\\[2ex]
-\frac{1}{2}\tilde{X}^{\mu}_{AB}&<(b^\ell)^{\mu}_{AB}<\frac{1}{2}X^{\mu}_{AB}\,,
\end{align}
\end{subequations}
for the vector coefficients as well as
\begin{subequations}
\label{eq:inferred-bounds-c}
\begin{align}
-\frac{1}{2}\tilde{X}^{\mu\nu}_{AB}&<(c^\ell)^{\mu\nu}_{AB}<\frac{1}{2}X^{\mu\nu}_{AB}\,, \displaybreak[0]\\[2ex]
-\frac{1}{2}\tilde{X}^{\mu\nu}_{AB}&<(d^\ell)^{\mu\nu}_{AB}<\frac{1}{2}X^{\mu\nu}_{AB}\,,
\end{align}
\end{subequations}
for the two-tensor coefficients. Note that the summary tables in ref.~\cite{Kostelecky:2008ts} list limits on the absolute values of the LV coefficients. Hence, as long as we take sensitivities from these particular tables, we will not have to make the distinction between $\tilde{X}^\mu$ and $X^\mu$, $\tilde{Y}^\mu$ and $Y^\mu$ as well as $\tilde{Z}^\mu$ and $Z^\mu$ (and analogously for the two-tensor-valued quantities). This also implies that the situation is identical for the coefficients $(a^\ell)^{\mu}_{AB}$ and $(b^\ell)^{\mu}_{AB}$ (see eq.~(\ref{eq:connection-constraints-1})) whenever we derive sensitivities from those listed in the summary tables. To infer the sensitivities on the charged-lepton sector given in tables~\ref{tab:constraints-a-inferred-charged-leptons}, \ref{tab:constraints-c-inferred-charged-leptons} of section~\ref{sec:connection-coefficients}, we employ the values of table~\ref{tab:original-neutrino-constraints} obtained from ref.~\cite{Kostelecky:2008ts}.

To compute the sensitivities on the minimal $(a_L)^{\mu}_{AA}$ coefficients, we discard eq.~(\ref{eq:two-sided-constraints-a-Y}) and only employ eq.~(\ref{eq:two-sided-constraints-a-Z}) to deduce
\begin{equation}
-2\tilde{Z}^{\mu}_{AA}<(a_L)^{\mu}_{AA}<2Z^{\mu}_{AA}\,.
\end{equation}
Using the latter inequalities and the values of table~\ref{tab:original-charged-lepton-constraints} we are able to deduce the sensitivities given in the first part of table~\ref{tab:constraints-inferred-neutrinos} in section~\ref{sec:connection-coefficients}. For the minimal $(c_L)^{\mu\nu}_{AA}$ coefficients, we choose the complete two-sided constraints of eqs.~(\ref{eq:two-sided-constraints-c-Y}), (\ref{eq:two-sided-constraints-c-Z}) for the dim-4 charged-lepton coefficients to infer new limits on LV in the neutrino sector. From eqs.~(\ref{eq:cR-cL-from-cell-dell}) in the main text we obtain
\begin{subequations}
\begin{equation}
-(\tilde{Y}+\tilde{Z})^{\mu\nu}_{AA}<(c_L)^{\mu\nu}_{AA}<(Y+Z)^{\mu\nu}_{AA}\,,
\end{equation}
which together with table~\ref{tab:original-charged-lepton-constraints} serves as a base to derive the sensitivities in the second part of table~\ref{tab:constraints-inferred-neutrinos}. Finally, all sensitivities stated in tables~\ref{tab:constraints-a-inferred-charged-leptons}, \ref{tab:constraints-c-inferred-charged-leptons}, and \ref{tab:constraints-inferred-neutrinos} are rounded to the leading significant digit, as the latter contains the essential information for such minuscule numbers. In principle,
\begin{equation}
-(\tilde{Y}+Z)^{\mu\nu}_{AA}<(c_R)^{\mu\nu}_{AA}<(Y+\tilde{Z})^{\mu\nu}_{AA}\,,
\end{equation}
\end{subequations}
could provide new bounds on right-handed charged leptons as a side effect, but we do not intend to take this possibility in consideration.

\section{Fit to neutrino time-of-flight data}
\label{sec:fit-neutrino-time-of-flight}

\begin{figure}
	\centering
	\includegraphics{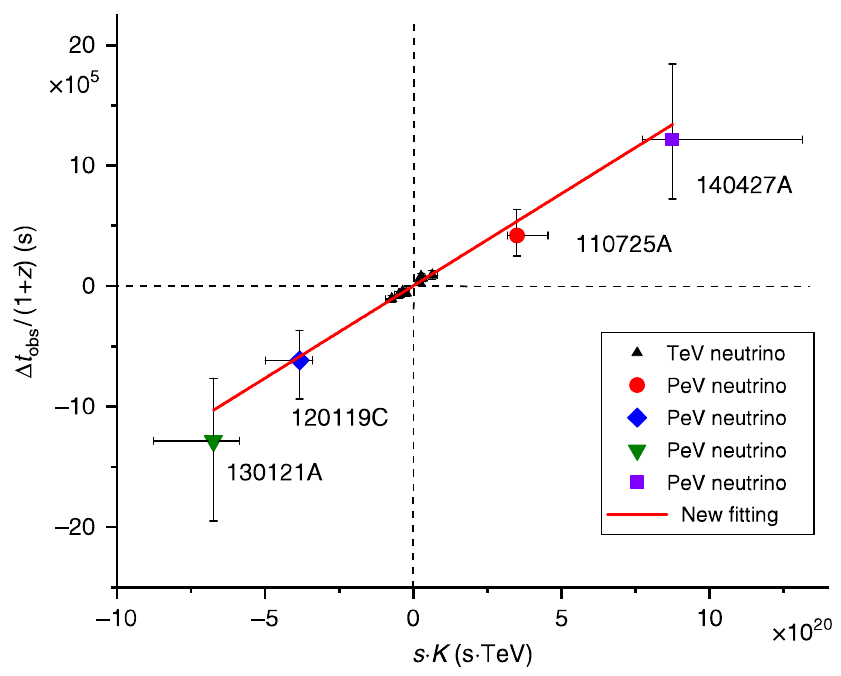}
	\caption{Linear fit presented in ref.~\cite{Huang:2018ham} to arrival time differences between GRB-neutrinos and \mbox{-photons} measured by IceCube. The vertical axis shows the observed arrival time difference corrected by the redshift factors $(1+z)^{-1}$ of the GRBs in consideration (to take into account the expansion of the universe). The horizontal axis displays the $K$ factor of eq.~(5) in ref.~\cite{Huang:2018ham} where a sign $s$ is taken into account for the late events. Both high-energy TeV events and four PeV events are shown in the plot.}
	\label{fig:arrival-time-difference-linear-fit}
\end{figure}

The linear fit obtained in ref.~\cite{Huang:2018ham} and leading to the value quoted in eq.~(\ref{eq:values-isotropic-coefficients}) of section~\ref{sec:consequences-neutrino-analyses} is reprinted in Fig.~\ref{fig:arrival-time-difference-linear-fit}. While this plot is intriguing, our arguments on $\mathit{SU}(2)_L$ gauge invariance developed in the main body of the text clearly demonstrate that LV in the neutrino sector cannot suitably explain why the data points exhibit the behavior found in Fig.~\ref{fig:arrival-time-difference-linear-fit}. A conventional reason for the goodness of the linear fit could be that the statistical spread of neutrino emission times (with respect to the emission times of photons) increases with neutrino energy.

Furthermore, clustering all delayed events in a single quadrant and the early events in the opposite one leads to a bias and automatically implies a straight line with positive slope. A plot of the absolute values along both axes is likely to reduce the significance of the finding. A related (though not equivalent) problem based on the clustering of events in opposite quadrants was one of the causes for the (false) announcement of the discovery of LV in polarization data of radio waves from quasars more than 20 years ago~\cite{Nodland:1997cc}. A subsequent reanalysis of the data (see, e.g.,~ref.~\cite{Carroll:1997tc}) showed that the polarization data did not exhibit a signal for LV.

Restrictions on the evolution of ultra-high-energy cosmic-ray sources actually disfavor active galactic nuclei and GRBs as being the sources of PeV neutrinos~\cite{Aartsen:2016ngq}. Even if the PeV-scale IceCube neutrinos did originate in GRBs, uncertainties on the differences between neutrino and photon emission times remain due to the model-dependent neutrino emission rates of GRBs~\cite{Guetta:2001cd,Meszaros:2001ms}.


\bibliographystyle{JHEP}
\bibliography{BIB}

%
%
%
%
%
%
%

\end{document}